\documentclass[draf]{aa}
\usepackage{natbib}
\usepackage{color}
\usepackage{graphicx}
\usepackage[varg]{txfonts}
\usepackage{hyperref}
\hypersetup{citecolor=black, urlcolor=blue, linkcolor=black, colorlinks=true}
\newcommand{\sect}[1]{Sect.\,\ref{#1}}

\newcommand{\fig}[1]{Fig.\,\ref{#1}}

\definecolor{orange}{rgb}{1,0.4,0.}
\definecolor{purple}{rgb}{1,0,1}
\graphicspath{{./}{figs-ps/}{figs/}}

\begin{document}

\title{A self-consistent 3D MHD model producing a solar blowout jet} 
\titlerunning{Blowout jet}

\author{Yajie Chen \inst{1}
          \and
          Hardi Peter \inst{1,2}
          \and
          Damien Przybylski  \inst{1}
          \and
          Lakshmi Pradeep Chitta  \inst{1}
          \and
          Sudip Mandal  \inst{1}
          }

\institute{
            Max-Planck Institute for Solar System Research (MPS),
            37077 G\"{o}ttingen, Germany
            \email{cheny@mps.mpg.de}
            \and
            Institute for Solar Physics (KIS), 
            Georges-K{\"o}hler-Allee 401A, 
            79110 Freiburg, Germany
          }

\date{Version: \today}

  \abstract
  {
  Solar blowout jets are a distinct subclass of ubiquitous extreme-ultraviolet (EUV) and X-ray coronal jets.
  }
  {
  Most existing models of blowout jets prescribe an initial magnetic field configurations and apply ad-hoc changes in the photosphere to trigger the jets.
  In contrast, we aim for a self-consistent magneto-convective description of the jet initiation.
  }
  {
  We employ a 3D radiation magnetohydrodynamic (MHD) model of a solar coronal hole region using the MURaM code. The computational domain extends from the upper convection zone to the lower corona.
  We synthesize the emission in the extreme UV and X-rays for a direct comparison to observations and examine the evolution of the magnetic field structure of the event.
  }
  {
  In the simulation a twisted flux tube forms self-consistently, emerges through the surface and interacts with the pre-existing open field.
  Initially the resulting jet is of the standard type with a narrow spire.
  The release of the twist into the open field causes a broadening of the jet spire turning the jet into a blowout type.
  At the same time this creates a fast heating front propagating at the local Alfv\'{e}n speed.
  The properties of the modeled jet closely match observations of blowout jets: a slow ($\sim$180\,km~s$^{-1}$) mass upflow and a fast ($\sim$500\,km~s$^{-1}$) propagating front form, the latter being a signature of the heating front.
  Also the timing of the jet with respect to the flux emergence and subsequent cancellation matches observations.
  }
  {
  The near-surface magneto-convection self-consistently generates a twisted flux tube that emerges through the photosphere.
  The tube then interacts with the pre-existing magnetic field by means of interchange reconnection. This transfers the twist to the open field region and produces a blowout jet that matches the main characteristics of this type of jet found in observations.
  }   

\maketitle

\section{Introduction\label{S:intro}}

Solar coronal jets are collimated plasma outflows that have been extensively studied, in particular since the soft X-ray observations of Yohkoh \citep[e.g.,][]{1992PASJ...44L.173S,1996PASJ...48..123S}.
They are often observed in extreme-ultraviolet (EUV) wavelengths \citep[e.g,][]{1999SoPh..190..167A,2008ApJ...680L..73P,2010ApJ...710.1806D,2012ApJ...759..144T} and sometimes have  chromospheric counterparts that are referred to as surges \citep[e.g.,][]{1995SoPh..156..245S,2003ApJ...584.1084C,2004ApJ...610.1136L,2008A&A...478..907C}.
They typically have lengths of tens of megameters, widths of several megameters, and lifetimes of minutes in EUV or X-ray observations \citep[e.g.,][]{2007PASJ...59S.771S}.
Some coronal jets extend to several solar radii and are observed in white-light coronagraph images \citep[e.g.,][]{1998ApJ...508..899W,2013RAA....13..253H,2015ApJ...806...11M}.
For recent reviews on coronal jets, we refer readers to \citet{2016SSRv..201....1R} and \citet{2021RSPSA.47700217S}.

\citet{2010ApJ...720..757M} suggested that coronal jets can be categorized into two classes, namely standard and blowout jets, based on their morphology. 
The spires of standard jets remain thin and narrow during their lifetime, while those of blowout jets broaden to the width of the jet bases and often present untwisting motions of helical structures \citep[e.g.,][]{2011ApJ...735L..43S,2012RAA....12..573C}.
The morphology of jets can change between standard and blowout types during their lifetimes \citep{2011ApJ...735L..18L,2022A&A...664A..28M}.
Blowout jets are often associated with eruptions of mini-filaments \citep[e.g.,][]{2013ApJ...769..134M,2014ApJ...783...11A,2015Natur.523..437S,2018MNRAS.476.1286J} and exhibit strong emission from cooler plasma observed in the 304 {\AA} \citep[e.g.,][]{2010ApJ...720..757M} and Ly$\alpha$ 1216 {\AA} passbands \citep{2023ApJ...944...19L}.
Blowout jets seem to have two velocity components: a bulk mass flow at $\sim$200 km~s$^{-1}$, and a faster component at $\sim$700 km~s$^{-1}$ generally seen in channels imaging hotter plasma (e.g., X-ray). This is interpreted as signatures of Alfv\'en waves propagating with the jets \citep{2007Sci...318.1580C,2023ApJ...944...19L}.
{Although the distinction between standard and blowout jets was originally introduced based on lower-resolution X-ray and EUV observations, higher-resolution EUV observations still reveal morphological differences, e.g., most small-scale standard jets remain narrow and lack untwisting motions and mini-filament eruptions \citep[e.g.,][]{2021ApJ...918L..20H,2023Sci...381..867C}.
Therefore, we use the term ‘blowout jet’ here to emphasize these specific characteristics, while recognizing that both types belong to the broader category of coronal jets.}

{Coronal jets are sometimes associated with flux emergence and/or cancellation \citep[e.g.,][]{1992PASJ...44L.173S,2008A&A...491..279C,2014PASJ...66S..12Y,2014SoPh..289.3313Y,2016ApJ...832L...7P,2018ApJ...853..189P,2021ApJ...909..133M,2021ApJ...918L..20H,2024ApJ...962L..38D}, which is often suggested as signatures of magnetic reconnection. 
Magnetic {extrapolations} before jet events often reveal 3D null points at their base \citep[e.g.,][]{2008ApJ...673L.211M,2012ApJ...746...19Z}, where breakout reconnection can occur and trigger coronal jets \citep[e.g.,][]{2018ApJ...854..155K,2019ApJ...873...93K,2024SoPh..299...88K}.}

To understand the formation mechanisms of coronal jets, various magnetohydrodynamic (MHD) simulations have been conducted.
Earlier 2D models demonstrated that standard jets can be produced by magnetic reconnection between emerging flux and {pre-existing} ambient magnetic field \citep[e.g.,][]{1995Natur.375...42Y,1996PASJ...48..353Y}.
Building on this, one group of models is constructed with twisted flux ropes emerging from below the surface, in which subsequent magnetic reconnection between the newly emerged twisted field lines and overlaying background magnetic field lines produces blowout jets \citep[e.g.,][]{2013ApJ...769L..21A, 2013ApJ...771...20M, 2014ApJ...789L..19F}.
Another group of models starts with an initial 3D null-point magnetic configuration.
Some of them drive the magnetic field at the surface by rotating motions to produce twisted structures in the closed loops and inject magnetic free energy into the system.
Later on instability-driven reconnection at the null point releases the stored energy and leads to blowout jets \citep[e.g.,][]{2009ApJ...691...61P,2015A&A...573A.130P,2017ApJ...834..123S,2017ApJ...834...62K,2017ApJ...837..123U}.
Some models with similar initial magnetic structures impose shearing motions at the surface to form flux ropes that later undergo {breakout} reconnection and trigger twisted jets \citep[e.g.,][]{2017Natur.544..452W,2018ApJ...852...98W,2018ApJ...864..165W}.
In addition, there are data-driven models that construct initial magnetic field configuration through extrapolation, reproducing the eruptive phase by manually adjusting the magnetic (or electric) field in the photosphere \citep[e.g.,][]{2015ApJ...801...83C,2019ApJ...875...10N} 
or by directly inserting flux ropes \citep[e.g.,][]{2022ApJ...938..150F,2025A&A...696L...2L}. 

In this study, we reproduce a blowout jet in a self-consistent way in 3D time-dependent radiation MHD simulations without prescribing any special magnetic field configurations for the jet or manually inserting flux tubes, and it naturally arises due to magneto-convection in our simulation.
We first compare the properties of the blowout jet in our model with those in observations and then investigate the underlying formation mechanisms driving the jet.

\section{MHD model and methods \label{S:model}}

The simulation was constructed from a snapshot of a quiet Sun model using the MURaM code \citep{MURaM,MURaM2017}, which {is described in detail in \citet{2025A&A...702L...4C}.}
The computational domain covers an area of 54$\times$54 Mm$^2$ in the horizontal directions with a grid size of $\sim$52.7 km.
The depth of the model is 20 Mm and the grid spacing is 20 km in the vertical direction.
The initial snapshot reaches to a height $\sim$500 km above the surface, and we added a uniform vertical component of magnetic field of 5\,G\ to replicate the flux imbalance in a coronal hole region.
After the simulation is saturated, we extended the model first to 6 Mm above the surface to form the transition region and then to 30 Mm to establish the corona.
After that we ran the simulation for an additional $\sim$10 hours, from which the magnetograms and the full 3D data cubes are collected with a cadence of $\sim$6 s and $\sim$6 min, respectively.
The model is maintained by a small-scale dynamo, and we did not prescribe flux emergence or any particular magnetic field structure except for the additional uniform 5 G added to the vertical component of the magnetic field at the very beginning. 

During the simulation run, a blowout jet is produced self-consistently. To better capture the jet dynamics, we restarted the simulation from the snapshot before the jet occurrence and collected output from our model at 10 s cadence for 20 minutes.
We define the time stamp of the first collected snapshot as $t=0$\,s.

To compare the model with observations, we synthesized emissions in three passbands, i.e., 304 {\AA} of the Atmospheric Imaging Assembly \citep[AIA,][]{2012SoPh..275...17L} onboard Solar Dynamics Observatory, 174 {\AA} of the Extreme Ultraviolet Imager \citep[EUI,][]{2020A&A...642A...8R} onboard Solar Orbiter, and Al-poly of the X-Ray Telescope \citep[XRT,][]{2007SoPh..243...63G} onboard Hinode.
Under the assumption of optically thin radiation, the emissivity at each voxel in a given passband is given by $n_e^2G_i(n_e,T)$, where $n_e$ and T are electron number density and temperature, and $G_i(n_e,T)$ is the contribution function, or kernel, of the passband.
We integrated the emissivity along a defined line of sight (LOS) to calculate the intensity maps. 

Although the emission in the AIA 304 {\AA} passband in real observations is not optically thin, our approach can provide a reasonable proxy for the emission from plasma at $\sim$10$^5$ K.
The synthesized EUI 174 {\AA} and XRT Al-poly images represent emission from plasma at temperatures at $\sim$10$^6$ K and $\sim$10$^{6.3}$ K, respectively.
To avoid the contamination of the emission from other bright structures along the LOS, we selected a small region that contains the blowout jet when we synthesized intensity maps.

\section{Results \label{S:results}}

\begin{figure*}[ht]
\sidecaption {\includegraphics[width=10cm]{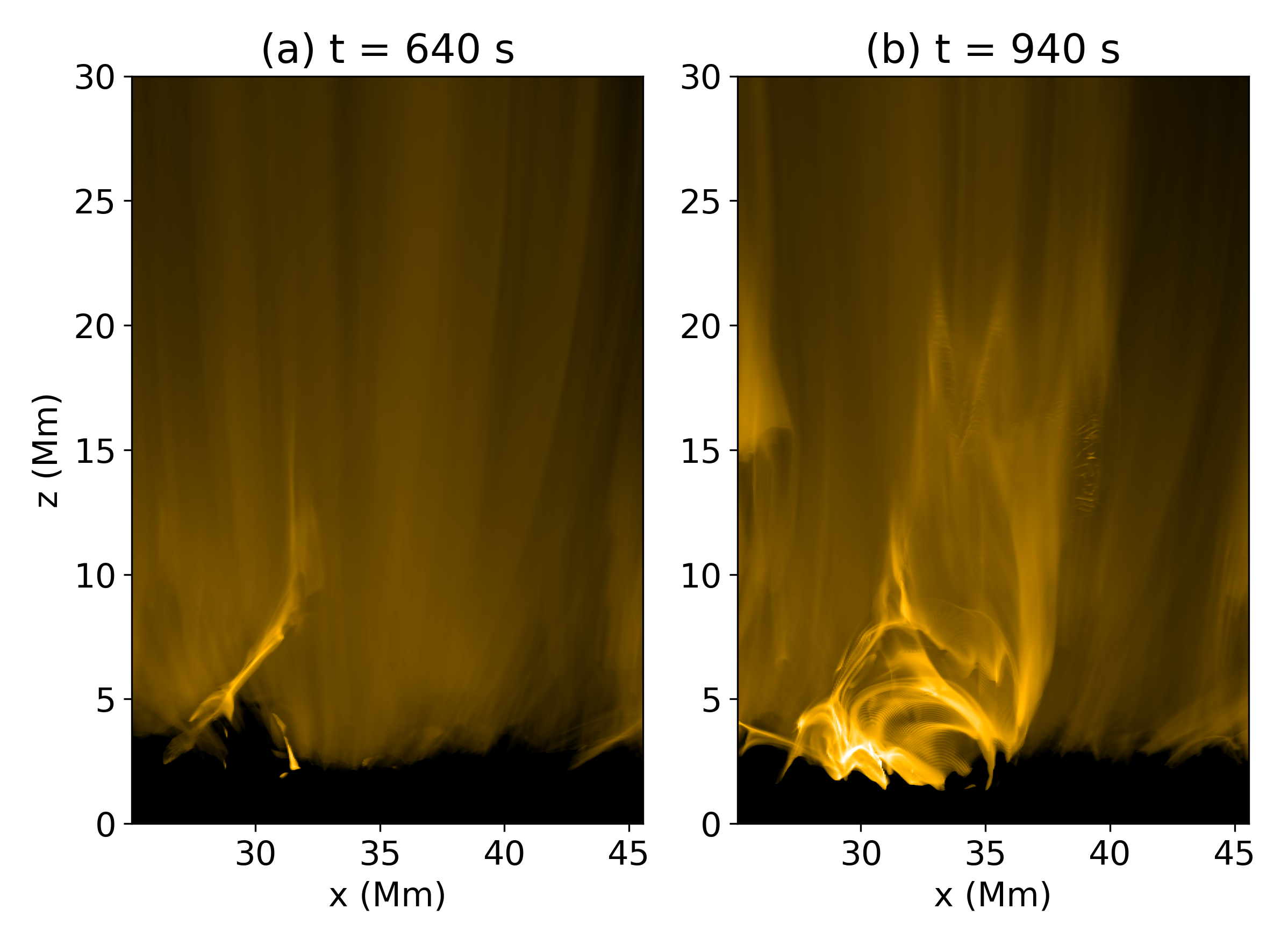}} 
\caption{Extreme UV images synthesized from a small region of the model showing two stages of the jet.
Panel (a) shows the standard-jet phase. Panel (b) depicts the situation only five minutes later when the jet transitioned into the blowout-jet phase.
Both images show the emission in the EUI 174 {\AA} passband integrated along the y-axis. This represents a view at or near the solar limb of plasma at temperatures around 1 MK.
See \sect{Sect:jet_dynamics}.
} 
\label{fig:2imgs}
\end{figure*}

\begin{figure*}[ht]
\centering {\includegraphics[width=15cm]{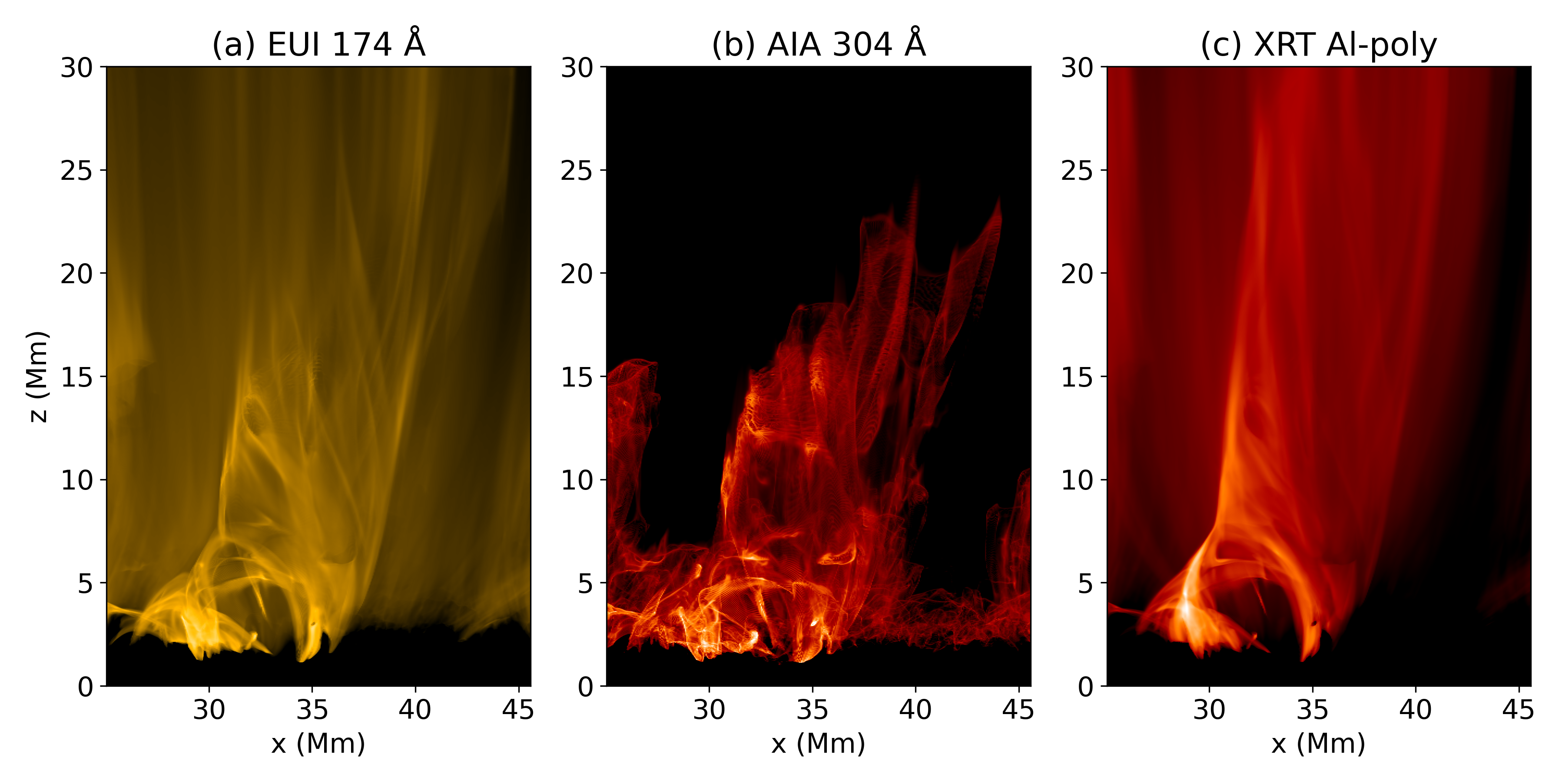}} 
\caption{Synthesized images of the blowout jet in different passbands. 
(a) Similar to \fig{fig:2imgs} but around time t${=}1000$\,s.
(b--c) Similar to (a), but for AIA 304 {\AA} and Hinode/XRT Al-poly passbands at the same time.
An animation is available online.
See \sect{Sect:jet_dynamics}.
} 
\label{fig:3passbands}
\end{figure*}

\subsection{Dynamics of the jet}\label{Sect:jet_dynamics}

We first present the temporal evolution of the jet in the EUI 174\,{\AA} passband in \fig{fig:2imgs}.
At the beginning, the jet exhibits a narrow spire and presents a Y-shaped morphology.
Several minutes later, the spire broadens to several megameters, comparable to the width of the jet base.
Thus the jet evolves from a standard type to a blowout type eruption, similar to some events in observations \citep{2011ApJ...735L..18L}.
The jet lasts for about 10 minutes and its emission signatures reach the top of the simulation domain, indicating that it can be traced to lengths of at least 30 Mm. These parameters are consistent with observational characteristics of blowout jets \citep{2016SSRv..201....1R}.

In addition, we present synthesized images in AIA 304 {\AA} and XRT Al-poly passbands during the blowout jet phase in \fig{fig:3passbands}.
The jet is clearly visible in all three passbands. In particular, it shows strong emission in the AIA 304 {\AA} passband, {with the 304 {\AA} emission originating from a solid core of cool plasma. It demonstrates} the presence of strong cool plasma within the jet.
This also matches with observations of blowout jets \citep{2010ApJ...720..757M}.

\begin{figure*}[ht]
\sidecaption {\includegraphics[width=10cm]{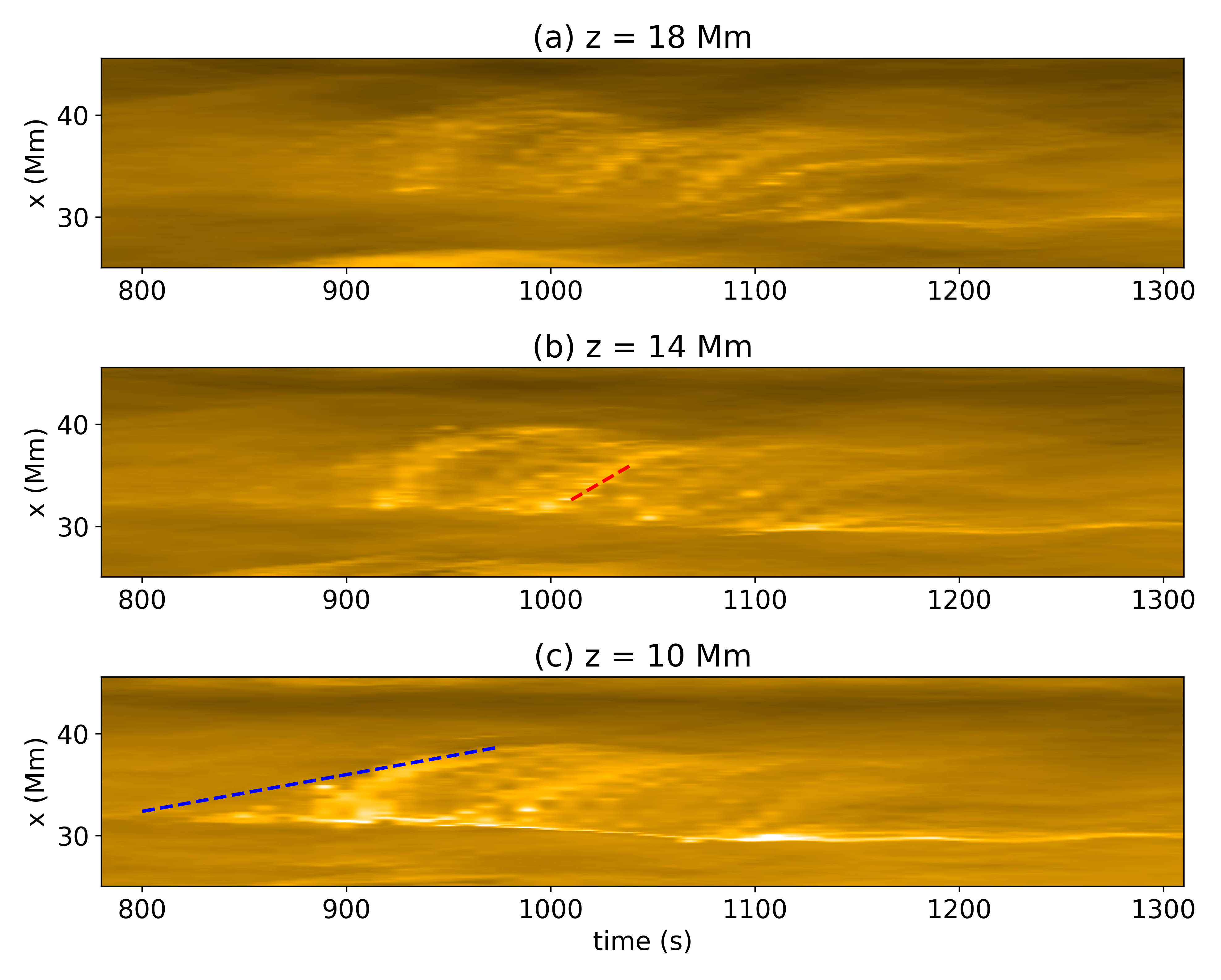}} 
\caption{Space-time diagrams across the jet spire at different heights.
These represent the temporal evolution at a horizontal line along the x-direction placed within the jet spire.
Panels (a) to (c) show the evolution at heights of 18, 14, and 10 Mm above the surface.
The sloped red {and} blue lines in panels (b) and (c) indicate apparent speeds of 116\,km\,s$^{-1}$ (red) and 36\,km\,s$^{-1}$ (blue).
See \sect{Sect:jet_dynamics} for details.
} 
\label{fig:st_plot_hor}
\end{figure*}

To further compare with observations, we placed three artificial slits at different heights and stacked the emission along these slits over time.
It is similar to placing a slit on imaging data in observations to construct a space-time diagram in observations.
The results are presented in \fig{fig:st_plot_hor}.
{During the blowout jet phase, the bright features in the space-time diagrams grow in size over time, as outlined by the blue dashed line in \fig{fig:st_plot_hor}(c), indicating radial expansion of the jet spire.
In the meantime, bright threads within the jet spire move laterally from one side of the jet to the other due to untwisting motions, producing bright stripes in the space-time diagrams. One example is outlined by the red dashed line in \fig{fig:st_plot_hor}(b).}
Both the radial expansion and untwisting patterns similar to those in observations \cite[e.g.,][]{2011ApJ...735L..43S,2012RAA....12..573C} are clearly seen in the diagrams at all heights.
Untwisting patterns mostly first occur at the lower heights and subsequently appear at higher heights, indicating that the untwisting motion propagates upward along the jet.
In addition, we calculated the radial expansion speed of the jet spire and the transverse (untwisting) speed of threads within the jet from the space–time diagrams. The radial expansion speed is approximately 36\,km\,s$^{-1}$, and the transverse component reaches up to 116\,km\,s$^{-1}$. 
These values are consistent with those reported by \citet{2011ApJ...735L..43S}.
{Here we selected one horizontal direction for comparison, and it is worth mentioning that choosing a different horizontal direction gives similar results.}

\begin{figure*}[ht]
\centering {\includegraphics[width=14cm]{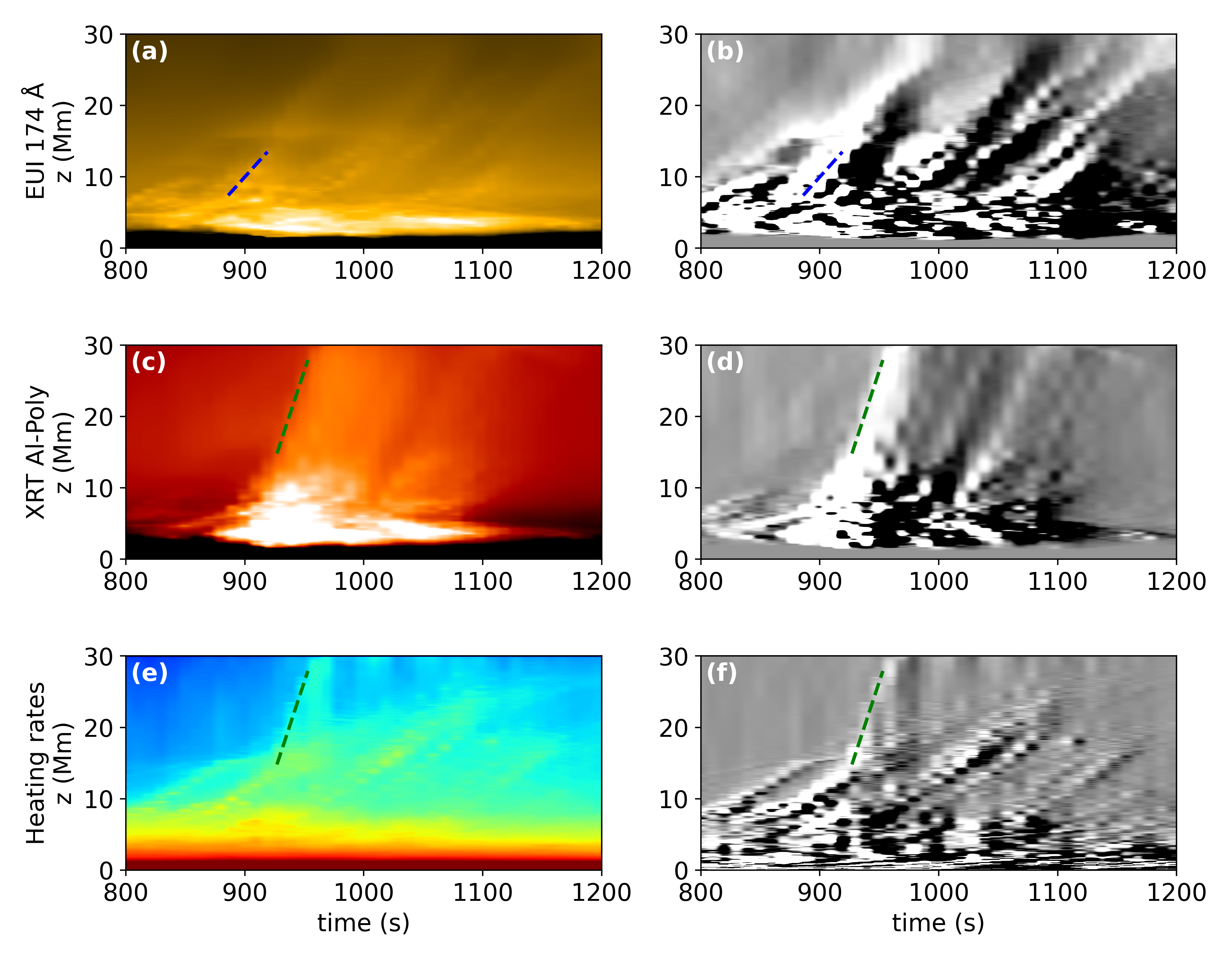}} 
\caption{
Space-time diagrams along the vertical direction.
These show the horizontally averaged vertical profile of the respective quantity as a function of time.
From top to bottom we display the evolution in EUI 174 {\AA} (a-b), Hinode/XRT (c-d) and in heating rate (e-f). The left column shows the original value of the emissivity (a,c) and heating rate (e), while the right {column} shows the respective running difference images (with time delay of 10\,s) simply to highlight the temporal changes. 
The dashed lines indicate the apparent speeds of 184\,km\,s$^{-1}$ (a-b) and 504\,km\,s$^{-1}$ (c-f).
See \sect{Sect:jet_dynamics}.
} 
\label{fig:st_plot_vert}
\end{figure*}

As a next step, we calculated the jet speeds in the vertical direction using synthesized EUI 174 {\AA} and XRT Al-poly images.
We integrated the emissivity of the two passbands horizontally for each snapshot and then stacked them to create the space-time diagrams shown in \fig{fig:st_plot_vert} (a) and (c). This shows the (horizontally averaged) propagation of the jet in the vertical direction as a function of time.
To highlight the propagation of the jet, we calculated running differences by subtracting the horizontally integrated emission of the previous snapshot ($\Delta$t$=$10 s) from each snapshot and stacking them into diagrams shown in \fig{fig:st_plot_vert} (b) and (d).
Based on these space-time diagrams, we estimated the jet speeds to be around 184~km~s$^{-1}$ in the EUI 174 {\AA} passband and about 504~km~s$^{-1}$ in the XRT Al-poly passband.
This is similar to the two-components of the jet speed as reported in observations  \citep{2023ApJ...944...19L}.

The lower speed in the EUI 174 {\AA} passband corresponds to bulk mass flows of the jet, while the higher speed in XRT Al-poly passband corresponds to Alfv\'en speed in the jet. 
However, the plasma velocity  in and around the jet {does} not exceed 400~km~s$^{-1}$ during the entire lifetime of the jet, hence, the higher apparent speed component cannot be explained by a bulk mass flow.
To investigate the origin of the high-speed  component, we present a space-time diagram of the heating rates (the sum of the Joule and viscous heating) along the vertical direction in \fig{fig:st_plot_vert}(e), similar to the plots for the EUI and XRT synthesized emission. 
This clearly shows an upward propagating feature similar as in the XRT emission in panel (c), with a speed matching that found in the synthesized XRT Al-poly images.
Therefore, the higher speed component seen in the hotter channels can be attributed to heating fronts.

\subsection{Magnetic origin of the jet} \label{Sect:magnetic_field}

\begin{figure*}[ht]
\sidecaption {\includegraphics[width=12cm]{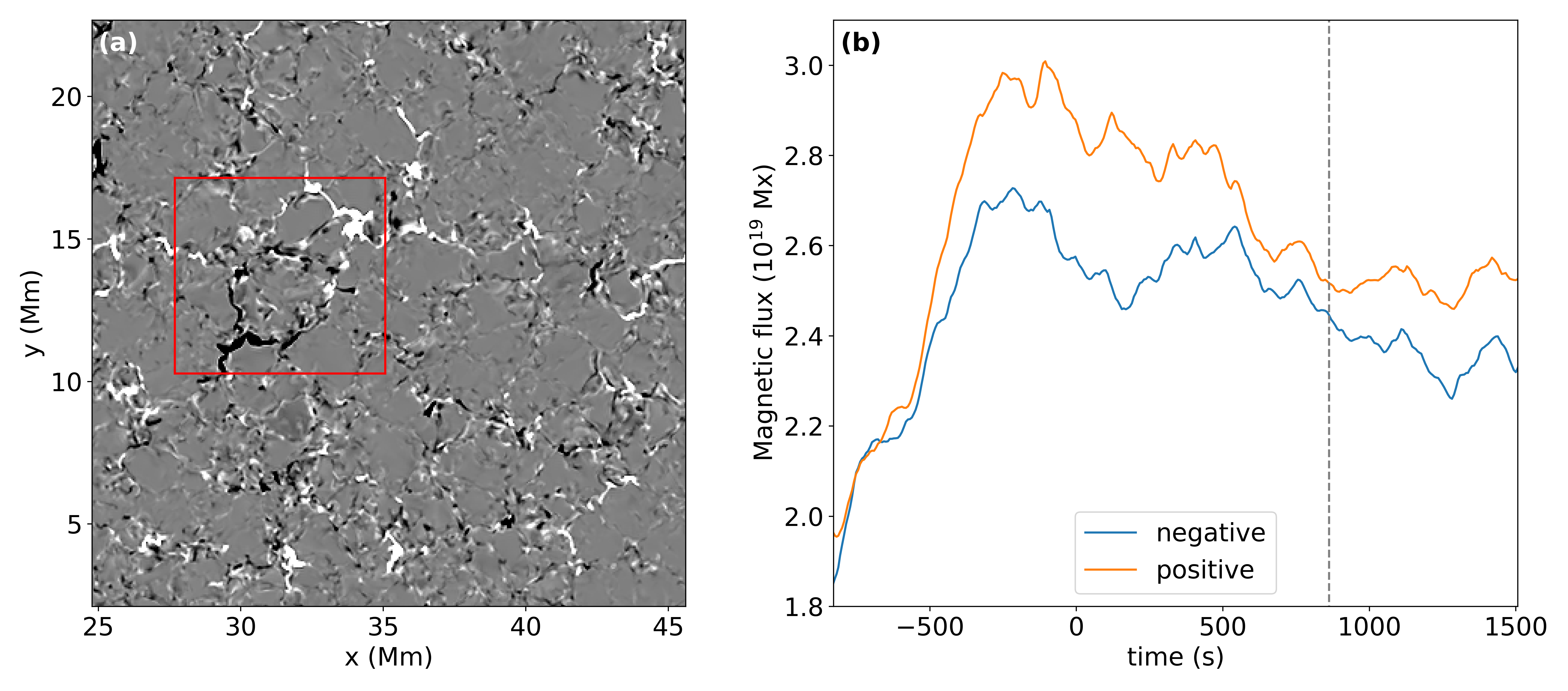}} 
\caption{
Temporal evolution of vertical component of magnetic field in the photosphere beneath the jet.
(a) Zoomed-in FOV of magnetogram beneath the jet.
(b) Orange and blue curves represent the sums of the absolute values of positive and negative polarities in the red box outlined in (a), respectively.
The vertical grey dashed line indicates the time of the magnetogram shown in (a).
An animation is available online.
See \sect{Sect:magnetic_field}.
} 
\label{fig:bz0}
\end{figure*}

We present the temporal evolution of the magnetic flux in the photosphere beneath the jet in \fig{fig:bz0}.
Flux emergence first occurs, accompanied by the formation of small-scale mixed polarities modulated by the Parker instability \citep{1966ApJ...145..811P}.
{The newly emerged opposite-polarity magnetic elements diverge and move apart by more than 5 Mm within about 10 minutes, indicating that the closed loops rise in height.}
Subsequently, flux cancellation among these newly emerged polarities occurs at the same locations.
The jet appears approximately 10 minutes after the magnetic flux reaches its maximum, coinciding with the decreasing trends in both positive and negative magnetic flux.
The delay between flux emergence and the appearance of the jet is because the newly emerged magnetic flux requires tens of minutes to rise to coronal heights and form a suitable magnetic configuration for reconnection, thereby producing signatures in coronal images.

The magnitudes of the changes in positive and negative magnetic flux are around 0.5$\times$10$^{19}$\,Mx, which is similar to the values found in observations \citep{2016ApJ...832L...7P}.
As a test, we slightly changed the location and size of the region used to calculate the magnetic flux (outlined as red box in \fig{fig:bz0}a) and found that it does not change our conclusions.

\begin{figure*}[ht]
\centering {\includegraphics[width=\textwidth]{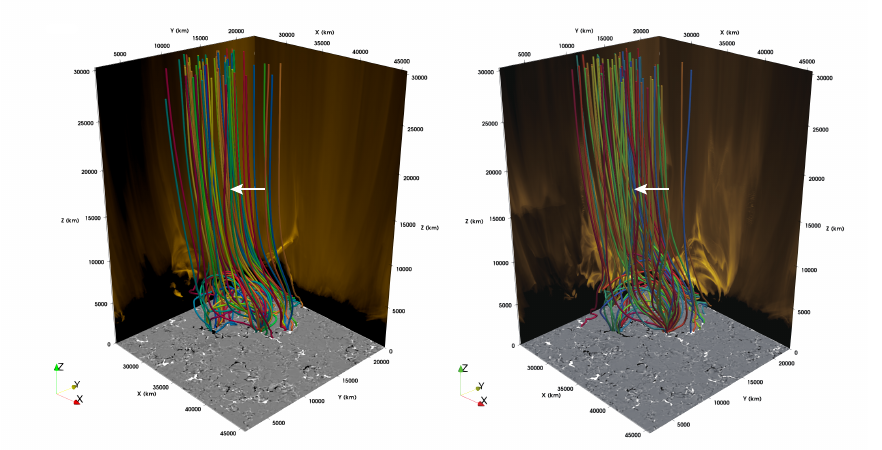}} 
\caption{Magnetic field structure surrounding the jet.
The two panels show the situation during the standard- (left) and blowout-jet (right) phases at the same times as in \fig{fig:2imgs}.
In each panel the bottom shows a zoom-in of the magnetogram at the respective time. The synthesized EUI 174 {\AA} images integrated along the x- and y- directions are depicted in the vertical slices of each 3D box.
The colored lines represent the magnetic field lines in and around the jet.
The white arrows point to field lines in the jet spire that become misaligned during the blowout jet phase (right), compared to the more aligned configuration during the standard phase (left). This indicates the untwisting motions that are also visible in the EUI 174 {\AA} images on the vertical slices in the animation showing the temporal {evolution}.
The animation is available online.
See \sect{Sect:magnetic_field}.
} 
\label{fig:field_line}
\end{figure*}

Flux emergence and/or cancellation are generally considered to be signatures of magnetic reconnection in the solar atmosphere \citep[e.g.,][]{2019Sci...366..890S}.
To examine the magnetic field topology in and around the blowout jet, we traced magnetic field lines from 80 seed points around the jet base in both backward and forward directions in space.
For simplicity, we fixed the locations of the seed points when tracing the field lines for different snapshots.

Magnetic field lines in two snapshots are presented in \fig{fig:field_line}, the left panel corresponds to the standard stage and the right panels shows the blowout stage.
Initially, twisted closed loops emerge into the corona and reconnect with pre-existing ambient background magnetic field lines.
{During this phase, the closed loops continuously rise and expand, increasing their heights before the onset of the jet.}
The jet originates at the interface between the closed and open field lines and appears as a standard jet, consistent with previous numerical studies \citep[e.g.,][]{1992PASJ...44L.173S,2006PASJ...58..423I}.
{The region of high current density expands as the jet base broadens, without showing a clear rotational drift similar to that reported by \citet{2010ApJ...714.1762P}.
As the current sheet expands, reconnection between the twisted closed loops and the open field occurs across a wide area.
The closed loops are opened and transfer their twist to the open field lines over an extended range at relatively low heights.
Because the field lines at higher heights are nearly straight, magnetic tension allows the twist to propagate upward, driving the blowout jet characterized by untwisting motions and the lateral expansion of the jet spire.}
One feature is that the field lines within the jet spire during the blowout jet phase {are} less well aligned compared to the standard jet phase as indicated by the white arrows in \fig{fig:field_line}.
This naturally explains the transition from the standard to the blowout stage.

{We have also applied the null-point detection method of \citet{2021ApJ...923..163P} to identify magnetic nulls within the jet and found a null point throughout the standard jet phase and the subsequent transition to the blowout jet phase, i.e. when the jet base expands. The null exhibits chaotic motions and jumps back and forth within an area of $\sim2\times2$Mm$^2$ in the xy-plane, without any clear rotational pattern. During the blowout jet phase the null point disappears, likely because the fan-spine topology is destroyed.}

\section{Discussion \label{S:discussions}}

A blowout jet is self-consistently generated in our model, reproducing many observational features associated with this type of jet.
By examining the magnetic field structure in and around the jet during its lifetime, we find that the jet is triggered by interchange reconnection between newly emerged twisted closed loops and neighboring open field lines.

In our model the twist in the emerging (magnetically closed) loops that emerge before the jet is built up in the convection zone prior to the flux emergence, because there are no clear rotational or shearing flows seen in the photosphere after the flux emergence.  
This behavior is quite different from previous numerical experiments where such surface motions play a key role in injecting the twist into the structure that produces the jet \citep[e.g.,][]{2015A&A...573A.130P,2017Natur.544..452W}.

The mechanism we find here, i.e., that a twisted magnetic structure emerges through the photosphere, is similar to previous MHD simulations of blowout jets that manually inserted a twisted flux rope to initiate the flux emergence and the jet \citep{2013ApJ...769L..21A,2013ApJ...769..134M,2014ApJ...789L..19F}.
However, the flux emergence in our model occurs self-consistently without the necessity to implant a flux rope beneath the surface by hand.
During the interchange magnetic reconnection with the pre-existing (open) magnetic field, twist is transferred from closed to open field lines, which drives the untwisting motion in the jet spire.

\citet{2023ApJ...944...19L} reported that blowout jets can exhibit two velocity components, and they interpreted the slower component as mass flow and the faster one as the untwisting of newly formed field lines during reconnection.
Similar high speed components in coronal jets were reported also in X-ray observations by \cite{2007Sci...318.1580C}, which the authors interpreted as signatures of Alfv\'en waves associated with jets.
Previous numerical studies have demonstrated that magnetic reconnection producing coronal jets can generate Alfvénic waves \citep[e.g.,][]{2014SoPh..289.3043L,2015A&A...573A.130P,2015ApJ...798L..10L,2022ApJ...941L..29W,2025ApJ...982L..25Y}. 
Nonlinear Alfvénic disturbances can steepen and develop shocks and current sheets, where energy can be dissipated, contributing to local heating \citep[e.g.,][]{1982SoPh...75...35H,1974PhFl...17.2215C}.

Our model successfully and self-consistently reproduces the two velocity components. The fast component ($\sim$500\,km\,s$^{-1}$) corresponds to heating fronts rather than real mass motion. This is supported through the speed of the fast component roughly matching the local Alfvén speed within the jet.
This is because magnetic perturbations triggered by reconnection travel along the field lines at the Alfv\'en speed, and the associated current sheets are apparently propagating along with these disturbances. 
The subsequent dissipation of the current sheets through Joule and viscous heating produces an apparent heating front.
As in any other 3D MHD model of the solar corona, the magnetic Reynolds and Prandtl numbers in our simulation are substantially different from those in the real corona. Still, the magnetic disturbances will cascade and steepen to small scales and dissipate energy at such fronts \citep{1982SoPh...75...35H,1974PhFl...17.2215C,2014ApJ...782...81V}.
The details of the dissipation might be different, but the overall physical picture remains robust.
Hence, our results are consistent with earlier studies where heating fronts have been invoked to interpret the high speeds seen in network jets observed in the transition region images \citep{2017ApJ...849L...7D,2014Sci...346A.315T}.

Although the blowout jet in our model resembles many observed features, we did not reproduce small-scale structures identified in some observations, such as mini-filament eruptions \citep[e.g.,][]{2015Natur.523..437S}.
{We did not find strong shearing motions along a polarity inversion line, and the emerging flux does not seem to form a filament channel. 
Therefore, a mini-filament is unlikely to form in our model.}
This mismatch might also be attributed to the treatment of the chromosphere in our model. Our current simulations do not include a proper treatment of the chromosphere \citep{2022A&A...664A..91P}, which could play an important role in the filament formation. 

In our model, flux cancellation at the jet base occurs mostly among newly emerged magnetic patches.
{We examined several cancellation events, and found that the regions of enhanced current density are mainly located in the photosphere and extend up to heights of
$\sim$500 km.
The associated downflows also originate from heights of a few hundreds kilometers. 
This indicates that many of the cancellation events in our model correspond to magnetic reconnection in the photosphere. 
Furthermore, we noticed that some canceling small-scale magnetic elements are initially connected to the major positive and negative magnetic concentrations, respectively. This implies that these reconnection events help to rearrange the magnetic footpoints connectivity and thereby modulate the fan-spine topology.}

However, in observations, some events appear associated with flux cancellation only, without clear signatures of flux emergence \citep[e.g.,][]{2014ApJ...783...11A}. Still, we consider it a possibility that in these earlier studies the emergence of new magnetic flux before the final cancellation during the jet event was simply missed in the observations. 
For some events studied by \citet{2016ApJ...832L...7P}, magnetic patches with opposite polarities approach each other before flux cancellation occurs, and flux cancellation typically lasts for several hours before the jet onset.
In our simulation, the jet is launched $\sim$10 minutes after flux cancellation begins.
Thus, the timescale for the accumulation of magnetic free energy in our model is significantly shorter than that in observations.

{
More generally, whether flux cancellation plays a key role in jet triggering remains debated. For example, \citet{2019ApJ...873...93K} found that only about 20\% of coronal-hole jets are associated with flux cancellation, and \citet{2021ApJ...912L..15T} showed that blowout jets can be triggered by external disturbances at coronal heights such as nearby eruptions. 
Thus, our model represents some coronal jets in which flux emergence and subsequent cancellation together drive the jet. In this scenario, flux cancellation is not necessarily the trigger itself but can arise naturally as a result of magnetic reconnection between the newly emerged and pre-existing magnetic fields.
}

Another observed feature in the jets missing in our model is the presence of small-scale blobs \citep[e.g.,][]{2017ApJ...851...67S,2019ApJ...873...93K,2022A&A...664A..28M,2023ApJ...944...19L,2023NatCo..14.2107C}.
The blobs are often interpreted as signatures of plasmoids triggered during magnetic reconnection \citep{2000mrp..book.....B}.
Previous high-resolution simulations of magnetic reconnection successfully reproduced plasmoids in jets \citep{2016ApJ...827....4W,2017ApJ...841...27N,2019A&A...628A...8P,2025A&A...702A.188N}.
Therefore, increasing the spatial resolution of our simulations might be necessary to capture plasmoids or blob-like features within the jet.

\section{Conclusions\label{S:conclusions}}

In this study, we presented a blowout jet that is self-consistently generated in a 3D radiation MHD model of a solar coronal hole region. 
To compare it with observations, we synthesized images in the AIA 304 {\AA}, EUI 174{\AA}, and XRT Al-poly passbands.
The width, length, and lifetime of the jet are consistent with observational properties.
Furthermore, the jet in our model exhibits strong emission from cool plasma in the synthesized AIA 304 {\AA} images and clear untwisting motions in the synthesized coronal images.
We identified two distinct speed components in the jet: a slower component ($\sim$180 km\,s$^{-1}$) corresponding to mass flows and a faster component ($\sim$500 km\,s$^{-1}$) associated with heating fronts.
In addition, flux cancellation occurring beneath the jet base was detected, {consistent with some observations}.
All these characteristics closely match with the observations.

Because of the good match to observations we assume that the underlying physics in our model provide a good model for what happens on the real Sun.
We examined the temporal evolution of magnetic field structures in and around the jet and found that the jet is triggered by magnetic interchange reconnection between newly emerged twisted closed loops and ambient open magnetic field lines. The twisted loops are generated self-consistently through the near-surface convection in the Sun and then emerge through the photosphere. This does not exclude the possibility that also surface motions might create this twisted magnetic structure that initiates the blowout jet. Still, our model demonstrates one clear way how such twisted structures form and emerge and then produce jets.
During the reconnection process, twist is transferred from closed loops to open magnetic field lines, driving magnetic disturbances associated with untwisting motions and heating fronts.
In conclusion, our model presents a self-consistent description on how the twisted pre-jet structures form and emerge, how they interact with the existing open magnetic field and how this finally produces the blowout jet at the base of the solar wind.

\begin{acknowledgements}
Y.C. acknowledges funding provided by the Alexander von Humboldt Foundation.
The work of Y.C., D.P., and S.M. was funded by the Federal Ministry for Economic Affairs and Climate Action (BMWK) through the German Space Agency at DLR based on a decision of the German Bundestag (Funding code: 50OU2201).
LPC gratefully acknowledges funding by the European Union (ERC, ORIGIN, 101039844).
We gratefully acknowledge the computational resources provided by the Cobra and Raven supercomputer systems of the Max Planck Computing and Data Facility (MPCDF) in Garching, Germany.
\end{acknowledgements}

\bibliography{refs}{}
\bibliographystyle{aa}

\end{document}